\documentstyle[12pt]{article}

\begin{document}
%\baselineskip=1.5\baselineskip

\renewcommand{\thefootnote}{\fnsymbol{footnote}}

\setcounter{equation}{0}
\setcounter{section}{0}
\renewcommand{\thesection}{\arabic{section}}
\renewcommand{\theequation}{\thesection.\arabic{equation}}

\pagestyle{plain}
\begin{titlepage}

\begin{flushright}
T93/127\\ hep-ph/9311315
\end{flushright}
\vfill
\begin{center}
{\large{  {Dimensional Continuation of Gauge-Invariant Quantities in \
Yang-Mills Theory\footnote{{\it Presented at the 3rd Thermal Fields \
Workshop at Banff, Canada,August 1993.}}}}}\par
\end{center}
\vskip 1.5cm
\centerline{Rajesh R. Parwani\footnote{email : parwani@wasa.saclay.
cea.fr}}
\vskip 0.5cm
\centerline{Service de Physique Th{\'e}orique, CE-Saclay}
\centerline{ 91191 Gif-sur-Yvette, France.}
\vskip 1.0cm
%\centerline{PACS 11.15Bt, 12.38Bx, 12.38Qk.}

\vskip 1.5 cm
\centerline{\bf Abstract}
\vskip 0.5cm
Consideration of some perturbatively calculated
 gauge-invariant expectation values of local noncomposite operators
in  pure Yang-Mills theory indicates that those expectation
values which are not dimension specific, and which are well defined
near $D=4$ dimensions, have some
{\it finite} limit as the evaluated expression is formally
extrapolated from $D=4$
 to $D=2$. If this finite limit is a  generic
feature for such quantities it would not only be quite remarkable
but also of some utility :
For example, one may
then use it as a convenient necessary condition when checking gauge
invariance. Some examples are discussed at nonzero temperature.

\vfill
\end{titlepage}

\section{Introduction}
Yang-Mills theory enables a  consistent description of
self-interacting spin one fields which form the backbone of both quantum
chromodynamics and the electroweak theory. As is well known, the underlying
fabric which holds Yang-Mills (YM) theory together is gauge-invariance, and
in perturbative quantum field theory this invariance, or redundancy, implies
 that only the $(D-2)$ transverse components of the
vector gauge field
describe propagating degrees of freedom in $D$ dimensions.
Indeed in any practical
calculation one is forced to break the classical gauge-invariance by
introducing a gauge-fixing term in the Lagrangian. However in the calculation
of gauge-invariant quantities, the final result should be insensitive to the
type of gauge-fixing and one often checks that this is indeed the case by
using an arbitrary gauge-fixing parameter. \\

Amusingly, precisely because Yang-Mills theory is a gauge theory, it
simplifies in a particular domain : that is for $D=2$ dimensions. Then there
are no physical particles and furthermore in the axial ($A_{1} =0$) gauge
the $D=2$ theory is manifestly noninteracting. A particular consequence of
this is that  gauge-invariant expectation values, of local operators,
in YM$_{2}$ are null since they are vanishing in the axial gauge. However
nonlocal operators, such as the Wilson loop (or perhaps even appropriately
constructed local but composite operators),
can have nonzero gauge-invariant expectation values even in a free theory
because such operators mimic external sources which couple to the gauge-fields.
For example, the Coulombic potential energy between two external static
charges in a theory with only photons is given by the large Euclidean
time limit of the expectation value of the Wilson loop. In what follows,
I will only consider local, noncomposite operators. \\

Suppose one could calculate some non-dimension-specific gauge-invariant
quantity to all orders in perturbation theory for an arbitrary
$D$ dimensions. Let's call this quantity $G(D)$. Then from the previous
discussion it follows that $G(D \to 2) = 0$ and one expects this limit to
be smooth. In practice, one can only compute  $G^{p}(D_{0};D)$ which is
a perturbative approximation to $G(D)$ valid for some values of $D$ near
$D =D_{0}$. The formal limit $G^{p}(D_{0}; D \to 2)$ then need not
vanish simply
because one is extrapolating beyond the region of validity of the calculation.
For example, the two-loop free energy of a gluon plasma is easily calculated
at and near $D=4$ but the expression already diverges as
$D \to 3^{+}$. Indeed, typically one finds $G^{p}(D_{0}=4; D \to 3) = \infty$,
reminding one that the boundary of the
region where the calculation is sensible has been approached
and that the further formal extrapolation
$D \to 2$ could, {\it a priori}, yield anything. \\

In Ref.[1], two infrared prescriptions were used in an attempt
to { \it enforce} the
limit $G^{p}(D_{0}=4; D \to 2) =0$. The
prescriptions appeared to be self-consistent in the examples considered but
it was not clear if this would always be the case. Here I would like to
expand on an observation made in the conclusion of Ref.[1] :
That without the
{\it ad hoc} infrared prescriptions, the formal limit $G^{p}(D_{0} =4;
D \to 2)$ was { \it finite} in the examples considered eventhough, as
discussed in the preceeding paragraph,
one might as well have expected a divergent result. Could it be that the
examples were hinting at a general result ?. Perhaps it is true that
perturbatively calculated gauge-invariant expectation values
in pure YM theory which are not
dimension specific, and which are
well-defined near $D=4$ dimensions, tend to a finite limit in the formal
extrapolation $D \to 2$. For conciseness, the words of this question will
be summarised as : ``$G^{p}(D_{0}=4; D \to 2) \neq \infty \ ? $''.
 By the phrase `well-defined' I exclude quantities which, for example, have
collinear infrared singularities that are aggravated as $D \to 2$ \cite{R}.\\

In the next section some examples are enumerated
with brief comments
and in the concluding section I explain why the question
``$G^{p}(D_{0}=4;D \to 2) \neq \infty \ ? $'' is `nontrivial'. As the
technical details and a more complete list of references for the examples
below has already been given \cite{R}, I will attempt not to repeat them.

\setcounter{equation}{0}
\section{Examples}
Here are some examples in $SU(N)$ YM theory at a nonzero temperature $T$.
The quoted expressions
represent sensible perturbative calculations for $3 < D \le 4$ and are
interpreted here by a formal extrapolation for $D < 3$. Note
the singularity as $D \to 3^{+}$ which signals the breakdown of the
naive perturbative calculation due to the appearance of infrared
singularities.\\

\noindent{\it 2.1 Free Energy}\\
The first three terms in the perturbative evaluation of the pressure (negative
of the free energy) are
\begin{equation}
P = P_0 + P_2 + P_3 \, .
\end{equation}
where
\begin{eqnarray}
P_{0} &=& (D-2) (N^{2}-1) \ T^{D} \ {\pi}^{-{D \over 2}} \ \Gamma(D/2) \
\zeta(D) \; , \label{IG2} \\
&& \nonumber \\
P_2 &=& -\left({{D-2} \over 2}\right)^2  g^{2} N(N^{2} -1) \ T^{(2D-4)} \
{\omega}^{2}(D) \  I^{2}(D) \; ,   \label{P23} \\
&& \nonumber \\
P_3 &=& { (N^2 -1) T \over 2 } \Gamma \left({1-D \over 2} \right) \
\left( {m_{el}^2 \over
4 \pi} \right)^{D-1 \over 2} \, , \label{P32}
\end{eqnarray}
with the the functions $\omega(D)$ and $I(D)$ defined by
\begin{eqnarray}
\omega(D) &=& \left[ 2^{(D-2)} \ \Gamma \left({D-1 \over 2}\right) \
\pi^{(D-1) \over 2}  \right]^{-1} \, ,\\
&& \nonumber \\
I(D) &=& \Gamma(D-2) \ \zeta(D-2).
\end{eqnarray}

The ideal gas result $P_0$ is actually valid for all $D \ge 2$ simply
because it is an exact statement of an explicitly solvable model : YM
theory at zero coupling. The
explicit $(D-2)$ factors in $P_0$ and $P_2$ come from the Lorentz algebra
while those in $P_3$ are only implicit in the definition of the
electric mass $m_{el}$ (see below).\\

\noindent{\it 2.2 Self Energy}\\
Consider the time-time component of the one-loop gluon self-energy in the
static limit, $ \Pi_{00}(k_0 =0 , \vec{k})$. In four dimensions, the first
two terms in its low momentum expansion give the electric mass squared
(at lowest order) and
a linear term in $|\vec{k}|T$. Though both terms are gauge-fixing independent,
let me add some clarification about the momentum dependent term : It has
been shown to be the same in general covariant gauges, the strict Coulomb gauge
\cite{EHKT,N} and also in a class of static gauges \cite{R2}. In the background
gauge its numerical value is different \cite{EHKT} but this is unsurprising
because Greens functions in nonabelian gauge theories do not necessarily
have the same physical (or gauge-invariant) interpretation in the presence
of a background field. Next, the linear term is susceptible to the effects
of resummation \cite{N} because it is a subleading (relative to order $g^2$)
quantity at low momentum. However if one focuses in the narrow window
$m_{el} \ll |\vec{k}| \ll T$ then resummation is unnecessary.
For $D$ near four, the two terms are
\begin{eqnarray}
m_{el}^2 &=& (D-2) \ g^2 \ N \ T^{(D-2)} \omega(D) \ \Gamma(D-1)
 \zeta(D-2) \, ,  \label{mD2}
\end{eqnarray}
and
\begin{equation}
T \left({D-2 \over 4} \right) { k^{(D-3)}  \over
{(4 \sqrt{\pi})^{(D-4)}}} { 1 \over \Gamma(D/2) \cos(\pi D/2)} \,
. \label{kT2}
\end{equation}

The $(D-2)$ factor in the momentum dependent term does {\it not} come from the
Lorentz algebra but rather from the integrals in dimensional reguralisation.
The function $\omega(D)$ is the same as defined above for the free energy.\\

\noindent{\it 2.3 Hard Thermal Loops}\\
The electric mass to lowest order is the static limit of a more general
gauge-invariant quantity : the leading $T^2$, momentum dependent, piece of the
one loop gluon self-energy. This ``hard thermal loop'' (HTL)
in the two-point function is in turn just one of an infinite set  of
gauge-invariant $n-$point ($n \ge 2$) HTL's \cite{BP}. These HTL's
may be summarised by a nonlocal generating functional, for which an
expression in $D$ dimensions may be written \cite{FT}.
The divergence of the expression in Ref.[6]
at $D=3$ is related to the fact that ``soft thermal loops''
are then no longer suppressed relative to the HTL's. (As an aside,
 the HTL  in the
photon self-energy in QED  is well-defined for all $2 < D \le 4$
because of Pauli repulsion).

\section{Conclusion}
The question of course is whether there are more examples to support the
suggestion {\bf {$ G^{p}(D_{0}=4; D \to 2) \neq \infty$ }}, or if one has to
impose more conditions in order to separate the ``bad apples'' from the
``chosen ones''.  If indeed some simple general statement could be made
along the lines indicated in the abstract,
then the result could be used as a
convenient check of gauge-invariance. This was the motivation for the
study in Ref.[1]. \\

It is noteworthy that a nondiverging limit is obtained
for $G^{p}(D_{0};D \to 2)$, at least for the examples considered so far.
If one were to take a similar limit for gauge {\it non-}invariant quantities
then very often the result is divergent : For example the $D \to 2$ limit
of Eq.(3) but excluding the ghost contribution,  or the same limit for
the one-loop
gluon self-energy (including ghosts) at zero temperature, both diverge. \\

Finally, for nonlocal operators such as the Wilson loop one can have
$ \infty > G(D=2) > 0$. One might ask if still, $ G^{p}(D_{0};D \to 2)
\neq \infty$ ?. For the heavy quark-antiquark potential at lowest order,
the answer is yes \cite{L}.\\

\noindent{ \bf Acknowledgements}\\
It is a pleasure to thank Randy Kobes and Gabor Kunstatter for a most
enjoyable, stimulating and instructive workshop.

\end{document}